\documentclass[prd,nofootinbib,preprint,superscriptaddress]{revtex4}

\usepackage{amsmath, amssymb, amsthm, graphicx, epsfig, fancyhdr,epsfig, slashed}

\usepackage{tikzsymbols}
\usepackage{natbib}
\usepackage{float}

\usepackage{tikz,xcolor,hyperref}

\definecolor{lime}{HTML}{A6CE39}
\DeclareRobustCommand{\orcidicon}{
	\begin{tikzpicture}
	\draw[lime, fill=lime] (0,0) 
	circle [radius=0.2] 
	node[white] {{\fontfamily{qag}\selectfont \tiny ID}};
	\draw[white, fill=white] (-0.0625,0.095) 
	circle [radius=0.007];
	\end{tikzpicture}
	\hspace{-2mm}
}

\foreach \x in {A, ..., Z}{\expandafter\xdef\csname orcid\x\endcsname{\noexpand\href{https://orcid.org/\csname orcidauthor\x\endcsname}
			{\noexpand\orcidicon}}
}

\newcommand{\be}{\begin{equation}}
\newcommand{\ee}{\end{equation}}
\newcommand{\bea}{\begin{eqnarray}}
\newcommand{\eea}{\end{eqnarray}}

\newcommand{\mt}{\mathtt{}}

\def\hhref#1{\href{http://arxiv.org/abs/#1}{arXiv:#1}} 

\begin{document}


\title{Confinement \& Renormalization Group Equations in String-inspired   Non-local Gauge Theories}

\author{Marco Frasca\orcidA{}}
\email{marcofrasca@mclink.it}
\affiliation{Rome, Italy}

\author{Anish Ghoshal\orcidB{}}
\email{anish.ghoshal@roma2.infn.it}
\affiliation{INFN - Sezione Roma “Tor Vergata”, Via della Ricerca Scientifica 1, 00133, Rome, Italy}
\affiliation{Institute of Theoretical Physics, Faculty of Physics, University of Warsaw,ul.  Pasteura 5, 02-093 Warsaw, Poland}

\author{Nobuchika Okada\orcidC{}}
\email{okadan@ua.edu}
\affiliation{Department of Physics and Astronomy, \\ University of Alabama, Tuscaloosa, AL 35487, USA}

\begin{abstract}
\textit{As an extension of the weak perturbation theory obtained in recent analysis on infinite-derivative non-local  non-Abelian gauge theories motivated from p-adic string field theory, and postulated as direction of UV-completion in 4-D Quantum Field Theory (QFT), here we investigate the confinement conditions and $\beta-$function in the strong coupling regime. We extend the confinement criterion, previously obtained by Kugo and Ojima for the local theory based on the Becchi-Rouet-Stora-Tyutin (BRST) invariance, to the non-local theory, by using a set of exact solutions of the corresponding local theory. We show that the infinite-derivatives which are active in the UV provides finite contributions also in the infrared (IR) limit and provide a proof of confinement, granted by the absence of the Landau pole. The main difference with the local case is that the IR fixed point is moved to infinity.
We also show that in the limit of the energy scale of non-locality $M \rightarrow \infty$ we reproduce the local theory results and see how asymptotic freedom is properly recovered.  }

\end{abstract}

\maketitle

\section{Introduction}



In the context of p-adic string field theory, recently, higher-derivative approaches to a UV-completion of QFT have become popular. Initially, they were proposed as possible UV-regularized theories \cite{Moffat:1990jj, Evens:1990wf, Tomboulis:1997gg, Moffat:2011an, Tomboulis:2015gfa,Kleppe:1991rv}. In this context, the infinite higher-derivative approach was motivated starting from string field theory \cite{sft1,sft2,sft3,padic1,padic2,padic3,Frampton-padic,marc,Tseytlin:1995uq,Siegel:2003vt,Calcagni:2013eua,Modesto:2011kw,Modesto:2012ga,Modesto:2015foa,Modesto:2017hzl} where attempts were made to address the divergence problem by generalizing the kinetic energy operators of the Standard Model (SM) to an infinite series of higher order derivatives suppressed by the scale of non-locality ($M$) at which the higher order derivatives come into the picture \cite{Krasnikov:1987yj,Biswas:2014yia}.

Such a theoretical construction naturally solves the SM vacuum instability problem  as the $\beta$-functions at the scale of non-locality vanish beyond $M$, without introducing any new degrees of freedom in the particle spectrum \cite{Ghoshal:2017egr}. 
They have been explicitly shown to be ghost-free 
\cite{Buoninfante:2018mre}, predicting conformal invariance in the UV, trans-planckian scale transmutation and dark matter phenomenology \cite{Ghoshal:2018gpq,Buoninfante:2018gce}.
This approach is a novel direction towards of UV-completion of 4D QFT, valid and perturbative up to infinite energy scales without Landau-pole problems \cite{Ghoshal:2017egr,Ghoshal:2018gpq}. Strong coupling regimes of the theory were studied in Refs. \cite{Frasca:2020jbe,Frasca:2021guf} , where it was shown that the mass gap obtained gets diluted in the UV due to non-local effects restoring conformal invariance in the UV. This leads to the fact that the infinite-derivative extension of the local theory behaves like a unique theory where conformal invariance at the classical level is maintained even at the quantum level; moreover even if scales are present in the infra-red (IR) of theory, the theory transcends into a scale-invariant one due to the presence of the higher-derivatives \footnote{Recently, it was shown that Lee Wick scalar theory with $N$ propagator poles, having $(N-1)$ Lee–Wick partners can be understood to flow to infinite-derivatives in the $N \rightarrow \infty $ asymptotic limit \cite{Boos:2021chb}.}.


Generally speaking, we employ the renormalization group equations (RGEs) to understand the relevance of UV fixed points for quantum field theories \cite{Wilson:1971bg,Wilson:1971dh}. For example, in quantum chromodynamics, the property known as the asymptotic freedom shows the reliability of the theory by the use of the standard perturbation theory \cite{Gross:1973id,Politzer:1973fx}. Asymptotic freedom is missing both in the U(1) and the scalar sector of the Standard Model, which is the so-called triviality problem, limiting the understanding of the behaviour of the theory at very large energies \cite{Callaway:1988ya}. When the fixed point corresponds to an interacting theory, we speak of asymptotic safety \cite{Weinberg:1980gg}.  It would be fine for theories, that are neither asymptotically free nor renormalizable, to have such a UV fixed point \cite{Litim:2011cp}. This idea was recently developed for quantum gravity \cite{Litim:2011cp,Litim:2006dx,Niedermaier:2006ns,Niedermaier:2006wt,
Percacci:2007sz,Litim:2008tt,Reuter:2012id}. Applications to the Standard Model also showed that such a possibility \cite{Litim:2014uca,Giudice:2014tma,Pelaggi:2017abg}. The authors studied the RGEs of non-local theories in Ref. \cite{Ghoshal:2017egr,Ghoshal:2020lfd} but only in the weak coupling regime. In this paper, we extend the same approach to the strong coupling regime by employing a novel technique which we will describe below.


Besides the RGE approach in QFT, deeper understanding of the confinement of quarks in the Standard Model (QCD sector) has eluded for many years, with some possible mechanism existing in the literature (see \cite{Kogut:2004su} and references there-in). Kugo and Ojima proposed a confinement condition from the BRST invariance based on charge annihilation \cite{Kugo:1977zq,Kugo:1979gm}. On a similar ground, Nishijima and collaborators \cite{Nishijima:1993fq,Nishijima:1995ie,Chaichian:2000sf,Chaichian:2005vt,Nishijima:2007ry} derived constraint on the amplitudes of unphysical states leading to color confinement merely as a consequence of the BRST invariance and asymptotic freedom of the QCD-theory.\footnote{In supersymmetric (SUSY) models, confinement is proven in certain conditions as a condensation of monopoles, similar to Type II superconductors \cite{Seiberg:1994aj,Seiberg:1994rs}. For a comparison of different confinement theories and their overlapping regions, see \cite{Chaichian:1999is}},\footnote{Studies by Gribov \cite{Gribov:1977wm} and Zwanziger \cite{Zwanziger:1989mf} suggested confinement in QCD with the gluon propagator running to zero as momenta go to zero and an enhanced ghost propagator running to infinity more rapidly than the free case in the same limit of momenta. These are supplemented via mass gap appearance from studies of the gluon and ghost propagators on the lattice \cite{Bogolubsky:2007ud,Cucchieri:2007md,Oliveira:2007px} and the spectrum \cite{Lucini:2004my,Chen:2005mg} in non-Abelian gauge theories without fermions. These results found theoretical basis in Refs. \cite{Cornwall:1981zr,Cornwall:2010bk,Dudal:2008sp,Frasca:2007uz,Frasca:2009yp,Frasca:2015yva} in terms of a closed form formula for the gluon propagator (see Ref. \cite{Deur:2016tte} for a review and other results for the running coupling in the infrared limit beside the gluon and ghost propagators).}  Confinement, in its simplest form, can be understood as the combined effect of a potential obtained from the Wilson loop of a Yang--Mills theory without fermions and the running coupling yielding a linearly incrementing potential, in agreement with lattice data \cite{Deur:2016bwq}. 

In this paper, we will apply the condition derived in \cite{Nishijima:1993fq,Nishijima:1995ie,Chaichian:2000sf,Chaichian:2005vt,Nishijima:2007ry} ( by reducing it to the case of the Kugo--Ojima criterion \cite{Kugo:1979gm} ) non-local non-Abelian gauge theories without fermions, as we begin with known exact solutions \cite{Frasca:2015yva,Frasca:2020jbe,Frasca:2021guf}. 
We will prove that non-local non-Abelian gauge theory with no fermions is confining in 4-D, besides having a mass gap coming from the derived correlation functions.This is purely because confinement arises due to the BRST invariance and the asymptotic freedom of the theory, as well as the existence of a mass gap. We show that the infinite-derivative operators defined in the UV yields finite contributions also in the IR-limit and provide a proof of confinement, granted by the absence of the Landau pole. 

\medskip

The paper is organized as follows: In Sec.\ref{Non_Abelian}, we introduce the non-local Yang-Mills theory for SU(N), starting by the well-known setting of the local theory. In Sec.\ref{BRST}, we extend the formalism of BRST invariance of the local theory to the non-local case and the Kugo-Ojima confinement criterion to the latter. In Sec.\ref{KO_conf}, we evaluate explicitly Kugo-Ojima criterion for the non-local case and, in Sec.\ref{beta}, we derive the beta function for the non-local theory that holds in the infrared limit. Finally, in Sec.\ref{conc}, some discussion and conclusions are presented.

\medskip

\section{Infinite-derivative Non-Abelian Gauge Theories: Review}
\label{Non_Abelian}

The Lagrangian for the SU(N) pure Yang-Mills theory, in the local case, takes the form 
\be
\mathcal{L}_{g}=-\frac{1}{4}  F^{a\mu\nu} F_{a\mu\nu}.
\ee
Repeated indexes imply summation both for space-time and group indexes. The field strength tensor is given by
\be
F_{\mu\nu}^a=\partial_{[\mu}A_{\nu]}^a -gf^{abc}A^b_{\mu}A^c_{\nu}     \ ,
\ee
with the group structure constants $f^{abc}$ and the dimensionless gauge coupling $g$. 
We extend the theory to the non-local case by following the approach given in Refs.~\cite{Ghoshal:2017egr,Ghoshal:2020lfd}. A common definition involving infinite-series of higher-derivatives yielded in literature \cite{Krasnikov:1987yj,Biswas:2014yia,Ghoshal:2017egr,Ghoshal:2018gpq,Ghoshal:2020lfd}:
\begin{equation}
    L_f = -\frac{1}{4}F^a_{\mu\nu}e^{-f(D^2)}F^{a\mu\nu}.
\end{equation}
Let us take
\begin{equation}
    f(D^2)=\frac{D^2}{M^2} ,
\end{equation}
where $D_\mu^{ab}=\partial_\mu \delta^{ab}-igA^c_{\mu}(T^{c})^{ab}$
is the covariant derivative in the adjoint representation. We have introduced a mass scale $M$ for the scale of non-locality. The scale is assumed to be large since non-local effects have not been experimentally observed so far. This implies that the variation in momentum scale of the $D^2$ is lower than $M^2$. For
\begin{equation}
    D^2=(\partial_\mu-igT^aA^a_\mu)^2=
    \partial^2-ig\partial^\mu\left(T^aA^a_\mu\right)-igT^aA^a_\mu
    \partial^\mu-g^2T^aT^bA^a_\mu A^{b\mu},
\end{equation}


one can apply the Backer-Campbell-Hausdorff formula (BCH), 
\begin{eqnarray}
    e^{-\frac{D^2}{M^2}}&=&e^{-\frac{\Box}{M^2}}e^{-\frac{1}{M^2}\left(-ig\partial^\mu\left(T^aA^a_\mu\right)-igT^aA^a_\mu\partial^\mu-g^2T^aT^bA^a_\mu A^{b\mu}\right)}\times \\
    &&e^{-\frac{1}{2M^4}\left[\Box,-ig\partial^\mu\left(T^aA^a_\mu\right)-igT^aA^a_\mu\partial^\mu-g^2T^aT^bA^a_\mu A^{b\mu}\right]}\times\ldots, \nonumber
\end{eqnarray}
and then the commutators are all higher orders with respect to the first two exponentials in the formula. Assuming that the field amplitudes are always negligible small with respect to the non-locality scale $M$, we can write
\begin{equation}
\label{eq:nlF2}
    L_f \sim -\frac{1}{4}F^a_{\mu\nu}e^{-f(\Box)}F^{a\mu\nu}.
\end{equation}
where $f(\Box) = f(\frac{\Box}{M})$. This affects gauge invariance but the effect turns out to be irrelevant, as the contribution arising from a change in the gauge potentials will produce higher order terms with respect to the non-locality scale. This means that we can assume Eqn.(\ref{eq:nlF2}) with negligible effect on our results. Such an argument is perfectly consistent with the computation reported in \cite{Tomboulis:1997gg} where the non-local factors in the propagators are evaluated using $\Box$ rather than $D^2$. The conventional Lagrangian is reproduced in the limit of $M \to \infty$. We take the metric convention with ${\rm diag}(+1,-1,-1,-1)$ to implement our procedure for UV completion upon the Wick rotation. \footnote{We will work in the Euclidean space with a certain relation between Minkowski and Euclidean metrics assuming the conclusions given in \cite{Pius:2016jsl,Briscese:2018oyx,Koshelev:2020fok,Koshelev:2021orf} which is done via analytical continuation. Practically at least in our computations, it is fine to use the standard Wick rotation.}

\medskip

\subsection*{Gauge Field Re-definition}

Let us define
\be
  {\hat A}_\mu^a=e^{-\frac{1}{2}f(\Box)}A_\mu^a.
\ee
The idea with this change of variables is to eliminate the exponential non-local factor from the kinetic term just as happens for scalar field theory. Indeed, our work will parallel the one already presented in \cite{Frasca:2020jbe}. With the re-definition, we will arrive at the Lagrangian
\begin{eqnarray}
\label{eq:nlYM1}
\mathcal{L}_{f}&=&\frac{1}{2}{\hat A}_\mu^a(\Box\eta^{\mu\nu}-\partial^\mu\partial^\nu){\hat A}_\nu^a \\
&&-\frac{g}{4}f^{abc}e^{-f(\Box)}\left[e^{\frac{1}{2}f(\Box)}\left(\partial_\mu {\hat A}^a_\nu-\partial_\nu {\hat A}_\mu^a\right)\left(e^{\frac{1}{2}f(\Box)}{\hat A}^{b\mu}
e^{\frac{1}{2}f(\Box)}{\hat A}^{c\nu}\right)\right] \nonumber \\
&&-\frac{g}{4}f^{abc}e^{-f(\Box)}\left[e^{\frac{1}{2}f(\Box)}{\hat A}^{b\mu}
e^{\frac{1}{2}f(\Box)}{\hat A}^{c\nu}e^{\frac{1}{2}f(\Box)}\left(\partial_\mu {\hat A}^a_\nu-\partial_\nu {\hat A}_\mu^a\right)\right] \nonumber \\
&&-\frac{g^2}{4}f^{abc}f^{cde}e^{-f(\Box)}\left[e^{\frac{1}{2}f(\Box)}{\hat A}^{b\mu}e^{\frac{1}{2}f(\Box)}{\hat A}^{c\nu}
e^{\frac{1}{2}f(\Box)}{\hat A}^{d}_\mu e^{\frac{1}{2}f(\Box)}{\hat A}^{e}_\nu\right] \nonumber \\
&&+j_\mu^ae^{\frac{1}{2}f(\Box)}{\hat A}^{a\mu},
\end{eqnarray}
where we added an arbitrary source term $j_\mu^a$ that will be useful in the following. This is similar to the way of a formulation of the non-local scalar field theory. The main difference is the multiplication of the interaction part by the non-local factor $e^{-f(\Box)}$.



The non-Abelian ghost and gauge-fixing Lagrangians are given by
\be
\mathcal{L}_{ghost}=-\bar{c}^ae^{-f(\Box)}(\partial^\mu D_{\mu}^{ab})c^b,
\label{NLghost}
\ee
and
\begin{equation}
\mathcal{L}_{g-f} = 
\frac{1}{2 \xi}{\hat A}_{\mu}^ae^{-f(\Box)}\partial^{\mu}\partial^{\nu}{\hat  A}_{\nu}^{a}, 
\label{NLgauge}
\end{equation}
where $\xi$ is the gauge fixing parameter. In order to have consistency with the standard gauge fixing procedure, we choose the entire function $e^{-f(\Box)}$. The standard QFT result can be obtained in the local limit of $M \to \infty$.

Similarly as we have done for the gauge field, the ghost field can be redefined as
\begin{equation}
    c^a=e^{-\frac{1}{2}f(\Box)}{\hat c}^a.
\end{equation}
This will yield
\begin{equation}
    \mathcal{L}_{ghost} =
    -{\bar{\hat c}^a}
    \partial^\mu\left(\partial_\mu \delta^{ab}-ige^{\frac{1}{2}f(\Box)}{\hat A}^c_{\mu}(T^{c})^{ab}\right)
    {\hat c}^b+{\bar\eta}^ae^{\frac{1}{2}f(\Box)}{\hat c}^a+e^{\frac{1}{2}f(\Box)}{\bar{\hat c}^a}\eta.
\end{equation}
Also in this case, we added arbitrary source terms $\eta^a$ and ${\bar\eta}^a$.



Finally, our Lagrangian is given by
\be
\label{eq:fullL}
\mathcal{L}=\mathcal{L}_{f}+\mathcal{L}_{g-f}+\mathcal{L}_{ghost}.
\ee

\medskip

\section{Confinement \& BRST Invariance}
\label{BRST}

We briefly describe the condition of confinement derived in \cite{Nishijima:1993fq,Nishijima:1995ie,Chaichian:2000sf,Chaichian:2005vt,Nishijima:2007ry}, and then we reduce it to the case of the Kugo--Ojima criterion \cite{Kugo:1979gm}, for non-local non-Abelian gauge theories without fermions, as we begin with known exact solutions  \cite{Frasca:2015yva,Frasca:2020jbe,Frasca:2021guf}.

\subsection{Local Theory}

Here we present the notations and the formalism as used in \cite{Chaichian:2018cyv}, for confinement conditions in local non-Abelian theories.

For the Yang-Mills field the following Lagrangian
\begin{equation}
\label{lagrangian}{\cal L}={\cal L}_{{inv}}+{\cal L}_{{gf}}+{\cal
L}_{{FP}}.
\end{equation}
Here, we have described $\cal L$ with
${\cal L}_{{inv}}$ for the classical gauge-invariant part, ${\cal L}_{{gf}}$ for the gauge-fixing terms and ${\cal L}_{{FP}}$ for the Faddeev--Popov (FP) ghost term proper to non-Abelian gauge theories:
\begin{eqnarray}
\label{eq:L2} 
{\cal L}_{{inv}}&=&-\frac{1}{4}F_{\mu\nu}\cdot
F^{\mu\nu}\,,\cr
%
{\cal L}_{{gf}}&=&\partial_\mu B\cdot A^\mu+\frac{1}{2}\xi B\cdot
B\,,\cr
{\cal L}_{{FP}}&=&i\partial_\mu \bar c\cdot D^\mu
c\,,\label{lagr_terms} 
\end{eqnarray}
%
where $\xi$ denotes the gauge parameter and $D_\mu$ is the covariant derivative given by 
%
\begin{eqnarray} 
D_\mu\ \psi&=&(\partial_\mu-igT\cdot A_\mu)\psi\,,\cr
 D_\mu\ c^a&=&\partial_\mu c^a+gf^{abc}A_\mu^b c^c\,.\label{cov_deriv}
\end{eqnarray}

As done in \cite{Kugo:1977zq}, we introduce the BRST transformations of a generic field $\chi$ using the BRST charges $Q_B$ and $\bar Q_B$ given by
\begin{eqnarray} 
\delta\,\chi=i[Q_B,\chi]_\mp,\ \ \ \bar\delta\,\chi=i[\bar
Q_B,\chi]_\mp\,,\\\label{brs_chi}
Q_B^2={\bar Q}_B^2=Q_B\bar Q_B+\bar Q_BQ_B=0\,.\label{brs_charge}
\end{eqnarray}
We will take the $-(+)$ sign in (\ref{brs_chi}) when $\chi$ is even (odd) in the ghost fields $c$ and $\bar c$. These are anti-commuting scalar fields.

The BRST transformations of the gauge field $A_\mu$ 
are generally defined by replacing the infinitesimal gauge function by the Faddeev-Popov (FP) ghost field $c$ or $\bar c$, in their respective infinitesimal gauge transformations
%
\bea
\delta A_\mu&=&D_\mu c\,,\cr
%
\bar\delta A_\mu&=&D_\mu \bar c\,,
\label{brs_transf} 
\eea
By imposing for the auxiliary fields $B$, $c$ and $\bar c$ 
\begin{equation} 
\delta{\cal L}=\bar\delta{\cal L}=0\,, 
\end{equation}
one gets
\begin{eqnarray} 
\delta\,B=0\,,\ \ \ \delta\,\bar c=i B\,,\ \ \
\delta\,c^a=-\frac{1}{2}gf^{abc} \,(c^b c^c)\,,\cr
\bar\delta\,\bar B=0\,,\ \ \ \bar\delta\,c=i \bar B\,,\ \ \
\bar\delta\,\bar c^a=-\frac{1}{2}gf^{abc}\,(\bar c^a\bar c^c)\,, 
\end{eqnarray}
with $\bar B$ defined by the following equation
\begin{equation} 
B^a+\bar B^a-igf^{abc}(c^b\bar c^c)=0\,.
\end{equation}

From Noether's theorem (up to a total divergence) one gets the conserved current
\begin{equation}
j_{\mu}=\sum_{\{\Phi\}}\frac{\partial \ L}{\partial (\partial_\mu \Phi)}\delta\Phi
=B^a(D_\mu c)^a -\partial_\mu B^a c^a+i\frac12{\rm g}f^{abc}\partial_\mu \bar c^a c^bc^c,
\end{equation}
with $\{\Phi\}$ being the set of all fields present in the Lagrangian. Therefore, the corresponding charge $Q_B$ is given by
\begin{equation}
Q_B=\int d^3x \left(B^a(D_0 c)^a -\dot B^a c^a+i\frac12{\rm g}f^{abc}\dot{\bar c}^a c^bc^c\right).
\end{equation}

Finally, the full Lagrangian will give
\begin{equation}
\delta({\cal L}_{{gf}}+{\cal L}_{{FP}})=\delta(-i\partial_\mu\bar c\cdot
A_\mu-\frac{i}{2}\xi\,\bar c \cdot B),
\end{equation}
confirming that
\begin{equation} 
\delta{\cal L}_{inv}=0\,. 
\end{equation}
%

From this Lagrangian we immediately obtain the equations of motion
\begin{equation}
   D^{\mu ab}F^b_{\mu\nu}+j^a_\nu=i\delta\bar\delta A^a_\nu.
\end{equation}
On the right-hand side, we see the contributions coming from the auxiliary fields.
These are massless particles at the tree level.
We also note that the $B$ field does not propagate. 
The consequences of this is that such fields will not give any contribution to the physical spectrum of the theory.
Besides, being $\partial^\nu(i\delta\bar\delta A_\nu)=0$, such a current is conserved. 
In order to evaluate such a contribution, we have to compute
\begin{equation}
\label{eq:corr}
\begin{array}{lll}
\langle i \delta \bar{\delta} A_{\mu} ^{a} (x), A_{\nu} ^{b} (y)
\rangle.
\end{array}
\end{equation} 

Referring to Kuog-Ojima formalism, we note that
\begin{equation}
   \delta \bar{\delta} A_{\mu}^{a}=-\{Q_B,\{\bar{Q}_B,A_{\mu} ^{a}\}\}.
\end{equation}
Then, because of $\langle 0|Q_B=Q_B|0\rangle=\bar{Q}_B|0\rangle=\langle 0|\bar{Q}_B=0$, one has
\begin{equation}
   \langle i \delta \bar{\delta} A_{\mu}^{a} (x),A_{\nu}^{b} (y) \rangle=\langle i  \bar{\delta} A_{\mu}^{a} (x),\delta A_{\nu}^{b} (y) \rangle=
	i\langle D_\mu\bar{c}^{a} (x),D_\nu c^{b} (y) \rangle.
\end{equation}
For this correlator, Kugo and Ojima showed  \cite{Kugo:1979gm} that
\begin{equation}
\label{eq:KOc}
\int d^dxe^{ipx}\langle D_\mu\bar{c}^{a} (x),D_\nu c^{b} (y) \rangle=\delta^{ab}
\left(\delta_{\mu\nu} - \frac{p_{\mu} p_{\nu}}{p^2-i\epsilon}\right) u(p^2)-\delta^{ab}\frac{p_{\mu} p_{\nu}}{p^2-i\epsilon},
\end{equation}
and the no-pole condition
yields here
\begin{equation} 
\label{eq:KO}
1+u(p^2=0)=0,
\end{equation}
which is the Kugo--Ojima condition for confinement granting that no massless pole appears in the spectrum of the theory. Indeed, this condition removes the massless term from Eqn.(\ref{eq:KOc})


The function $u(p^2)$ was computed explicitly for non-Abelain gauge theories in \cite{Chaichian:2018cyv}, and the confinement was proven for such theories; moreover it also yields an exact beta function for the theory including the strong coupling regimes. Next, we will extend this computation to the non-local case.

\medskip

\subsection{Non-local Theory}


The complete BRST-invariant infinite-derivative gauge theory in the quantized action is of the form \cite{Ghoshal:2020lfd}: 
\be \label{F1}
\mathcal{L}_{\mt{inv}} = -\frac{1}{4} (F_{\mu\nu}^a  e^{-f(D^2)} (F^a) ^{\mu\nu}) + \frac{\xi}{2} (B^a) ^2 + B^a \partial ^{\mu} A_{\mu}^a + \bar{c} ^a (-\partial ^{\mu} e^{-\frac{1}{2}f(\Box)} D_{\mu}^{ac})c^c,
\ee
where $\xi$ is the gauge fixing parameter, $B$ is the auxiliary field, and $c$ and $\bar{c}$ are the ghost and anti-ghost fields, respectively.
The BRST transformations for non-Abelian gauge theories express a residual symmetry
of the effective action which remains after the original gauge invariance has been
broken by the addition of the gauge-fixing and ghost action terms. 
Our BRST transformations are modified in the following way:

\begin{eqnarray}
\label{eq:brst1} 
\delta A_\mu&=&D_\mu c\,,\cr
\bar\delta A_\mu&=&D_\mu \bar c\,,\label{brs_transf} 
\end{eqnarray}
and
\begin{eqnarray} 
\label{eq:brst2}
\delta\,B=0\,,\ \ \ \delta\,\bar c=i e^{\frac{1}{2}f(\Box)}B\,,\ \ \
\delta\,c^a=-\frac{1}{2}gf^{abc} \,(c^b c^c)\,,\cr
\bar\delta\,\bar B=0\,,\ \ \ \bar\delta\,c=ie^{\frac{1}{2}f(\Box)}\bar B\,,\ \ \
\bar\delta\,\bar c^a=-\frac{1}{2}gf^{abc}\,(\bar c^b \bar c^c)\,.
\end{eqnarray}
We show the BRST-invariance of 
$S _ {\mt{inv} }$ by noting that
the BRST transformation of the gauge field is just a gauge transformation of $A_{\mu }$ generated
by $c_{a}$ or ${\bar c}_{a}$
. Therefore, any gauge-invariant functionals of $F_{\mu \nu}$, 
like the first term in Eqn.~(\ref{F1}) gives
$\delta(-\frac{1}{4} (F_{\mu\nu}^a  e^{-f(D^2)} (F ^{\mu\nu})^a)) = 0.$
The second term in Eqn.~(\ref{F1}) gives $\delta(\frac{\xi}{2} (B^a) ^2) = 0$ from Eqn.~(\ref{eq:brst2}).
For the third term in Eqn.~(\ref{F1}), the transformation of A$_{\mu} ^a$ cancels the transformation of $\bar{c}$ in the last term due to Eqs.~(\ref{eq:brst1}),  leaving us with
\be \label{F2}
\delta(D^{ac} _{\mu} c^c) = D^{ac} _{\mu} \delta c^c + g f^{a b c} \delta A ^b_{\mu} c^c,
\ee
 which is is equal to 0, using the \textit{Jacobi identity} (see Ref.~\cite{Peskin:1995ev}).
The transformation of $c^{\sigma}$ is nilpotent,
\be
\delta(\partial_{\mu} c^{a}c^{b})=0\,,
\ee
while the transformation of $A^{\mu}$ is also nilpotent,
\be
\delta((D_{b}^{\mu})^a c^{b}) = 0\,.
\ee
Hence, the action in Eqn.~(\ref{F1}) is BRST-invariant.
Noting the fact that the only part of the ghost action which varies under the BRST transformations is that of  the anti-ghost ($\bar{c} _{a}$), the central idea behind our proof of BRST-invariance is that we have chosen the BRST variation of the anti-ghost ($\bar{c} _{a}$) (see Eqs.~(\ref{eq:brst2})) to cancel the variation of the gauge-fixing term.


It is not difficult to see that, in the limit of the non-local mass $M\rightarrow\infty$, the BRST transformations given in Eqn.(\ref{eq:brst1})-(\ref{eq:brst2}) become identical to those of the local case. Formally, the confinement condition of Eqn.(\ref{eq:KO}) remains untouched as the effects of the non-locality, if present, are kept into the $u$ function.

\bigskip

\section{Condition of Confinement in Non-local theory}
\label{KO_conf}


 In this section, we derive the confinement for the non-local theory, following Ref.~\cite{Chaichian:2018cyv}. See Appendix C for a brief review of this technique.

From the action (\ref{F1}), we derive the equations of motion,
\be
e^{-f(D^2)}D^\mu F^a_{\mu\nu}+j_\nu^a=i\delta{\bar\delta}A_\nu^a.
\ee
The LHS can be evaluated as already done for the local case, and we write down
\be
\int d^4x e^{ipx}\langle D_\mu{\bar c}^a(x),D_\nu c^b(0)\rangle=\delta_{ab}\left(\delta_{\mu\nu}-\dfrac{p_\mu p_\nu}{p^2}\right)u(p^2)-\delta_{ab}\frac{p_\mu p_\nu}{p^2}e^{f(-p^2)}.
\ee
Indeed, this is the most general form for the given correlation function but, for the massless contribution, we have also to take into account the contribution of the non-locality. The interesting part here is that all the non-local contributions enters into the definition of the function $u$.
These non-localities arise from the two-point functions of the non-local theory but also that fluctuations from UV can yield a significant contribution to confinement as they are summed up in the integral where they cannot be neglected. Then, the confinement condition is again
\be
1+u(p^2=0)=0 .
\ee
To evaluate this equation, we will have for the two-point function
\cite{Frasca:2021guf}
\be
\label{eq:G2s}
G_2(p)=\frac{e^{\frac{1}{2}f(-p^2)}}{p^2+\Delta m^2e^{\frac{1}{2}f(-p^2)}}\frac{1}{1-\Pi(p)}.
\ee


In the local limit, $M\rightarrow\infty$, Eqn.(\ref{eq:G2s}) reduces to a Yukawa form that yields a fair approximation to the exact local propagator obtained in \cite{Frasca:2015yva}. In the non-local case, one has the mass gap
\be
\label{eq:mg}
\Delta m^2=
\mu^2
 \left(18Ng^2\right)^\frac{1}{2}\frac{4\pi^2}{K^2(i)}
 \frac{e^{-\pi}}{(1+e^{-\pi})^2}e^{f\left(-\frac{\pi^2}{4K^2(i)}p^2\right)}
 +\delta m^2.
\ee
This must be completed by the gap equation
\be
\label{eq:mgge}
\delta m^2=2Ng^2G_2(0)=2Ng^2\int\frac{d^4p}{(2\pi)^4}G_2(p).
\ee
The function $\Pi(k)$ can be neglected as also the shift $\delta m^2$ as a first approximation. Similarly, for the ghost one has
\be
K_2(p)=-\frac{1}{p^2}e^{\frac{1}{2}f(-p^2)}.
\ee
The confinement condition can be written as \cite{Chaichian:2018cyv}
\begin{eqnarray}
\int d^4xe^{ipx}\langle D_\mu\bar{c}^{a} (x),D_\nu c^{b} (0) \rangle&=&-\delta^{ab}\frac{p_\mu p_\nu}{k^2}\\
&+&\frac{(N^2-1)^2}{2N}g^2\delta^{ab}
\left(\delta_{\mu\nu} - \frac{p_{\mu} p_{\nu}}{p^{2}}\right)
\int\frac{d^4p'}{(2\pi)^4}K_2(p-p')G_2(p').\nonumber
\end{eqnarray}
This will yield for the confinement condition
\be
\label{eq:NL}
u(0)=-\frac{(N^2-1)^2}{2N}g^2
\int\frac{d^4p}{(2\pi)^4}\frac{1}{p^2}
\frac{e^{f(-p^2)}}{p^2+\Delta m^2e^{\frac{1}{2}f(-p^2)}}.
\ee
From this integral, we see explicitly how the dependence on the non-locality scale does appear. We will get contributions from the far UV that cannot be neglected as they sum up. Indeed, the non-local scale $M$ imposes a truncation in the spectrum and this turns out to be equivalent to computation into infrared having singular UV behavior that cannot be neglected and a cut-off dependence pops out, like for Nambu-Jona-Lasinio models of low-energy QCD \cite{Klevansky:1992qe}. On the other way round, in local QCD, asymptotic freedom entails a dimensional transmutation with the appearance of a mass scale in the theory \cite{Gross:1973id,Politzer:1973fx}. Indeed, in the local limit, $M\rightarrow\infty$, we are able to recover an approximate beta function by solving the integral
\be
\label{eq:LL}
u_{LL}(0)=-\frac{(N^2-1)^2}{2N}g^2
\int\frac{d^4p}{(2\pi)^4}\frac{1}{p^2}
\frac{1}{p^2+\Delta m^2}.
\ee
This is a fair approximation to the result presented in \cite{Chaichian:2018cyv} for the local theory. We have here a single Yukawa propagator while, in the local case, we have an infinite sum of such propagators. 

Turning back to Eqn.(\ref{eq:NL}), we emphasize that this result is just an approximate one but holds when the non-local effects are seen to modify significantly the spectrum of the theory. 
The integral can be computed only when the entire function is fully specified, which we assume
\be
f(-p^2)=e^{-\frac{p^2}{M^2}}.
\ee
Therefore, the above integral can be rewritten in the form
\be
u(0)=-\frac{(N^2-1)^2}{(2\pi)^dN}g^2\frac{2\pi^{\frac{d}{2}}}{\Gamma(d/2)}\frac{M^{d-2}}{\Delta m^2}\left[\frac{1}{2}\Gamma\left(\frac{d}{2}-1\right)
-f\left(\frac{\Delta m^2}{M^2}\right)\right],
\ee
where
\be
f(z)=\int_0^\infty dx x^{d-1}\frac{e^{-x^2}}{x^2+ze^{-\frac{x^2}{2}}}.
\ee
For $z=\Delta m^2/M^2\ll 1$, we have the expansion
\be
f(z)=\frac{1}{2}\Gamma\left(\frac{d}{2}-1\right)
-2^{\frac{d}{2}-3}3^{2-\frac{d}{2}}\Gamma\left(\frac{d}{2}-2\right)z
+2^{2-\frac{d}{2}}\Gamma\left(\frac{d}{2}-3\right)z^2+O(z^3).
\ee
It is interesting to point out that, using Eqn.(\ref{eq:mg}), the development parameter $z$ will be essentially given by the ratio $\mu^2/M^2$, with $\mu$ being the characteristic scale fixing the ground state of the theory and arising from as an integration constant. This just says to us that the infrared limit is the reliable one for our computation.

We can introduce the $\epsilon$ parameter as $d=4-\epsilon$ and we have
\be
u(0)=-\frac{(N^2-1)^2}{(2\pi)^{4-\epsilon}N}g^2\frac{2\pi^{2-\frac{\epsilon}{2}}}{\Gamma(2-\epsilon/2)}M^{-\epsilon}z^{-1}
\left[2^{-\frac{\epsilon}{2}-1}3^{\frac{\epsilon}{3}}\Gamma\left(-\frac{\epsilon}{2}\right)z
-2^{-\frac{\epsilon}{2}}\Gamma\left(-1-\frac{\epsilon}{2}\right)z^2+O\left(z^3\right)\right]
\ee
Finally, we expand in $\epsilon$ to get
\be
u(0)=-\frac{(N^2-1)^2}{8\pi^2N}g^2z^{-1}
\left[\left(-\frac{1}{\epsilon}-\frac{\gamma}{2}\right)z
-\left(\frac{2}{\epsilon}+\gamma-1\right)z^2+O\left(z^3\right)\right]+O(\epsilon).
\ee
We observe that the integral has a divergent part given by
\be
u_\infty(0)=\frac{(N^2-1)^2}{8\pi^2N}g^2\frac{1}{\epsilon}\left(1+2z+O(z^2)\right)
\ee
that we reabsorb through a redefinition of the coupling, being the theory renormalizable. Then, we write
\be
\label{eq:u0}
u(0)=\frac{(N^2-1)^2}{2\pi N}\alpha_s
\left[\frac{\gamma}{2}
-(1-\gamma)z+O\left(z^2\right)\right]+O(\epsilon),
\ee
where $\alpha_s=g^2/4\pi$. Thus, non-local Kugo-Ojima confinement condition yields the running coupling equation
\be
\label{eq:coup}
\frac{(N^2-1)^2}{2\pi N^2}\alpha_s
\left\{\frac{\gamma}{2}
-(1-\gamma)\frac{\Delta m^2}{M^2}+O\left[\left(\frac{\Delta m^2}{M^2}\right)^2\right]\right\}=-1,
\ee
where we have redefined $\alpha_s\rightarrow N\alpha_s$ to introduce the 't Hooft coupling. This is a confinement condition that depends on $\alpha_s$ in a highly non-trivial way. Working on shell, we have from Eqn.(\ref{eq:mg}),

\be
\Delta m^2=\mu^2\alpha_s^\frac{1}{2}\eta_0
e^{-\eta_1\mu^2\alpha_s^\frac{1}{2}/M^2}
\ee
with numerical constants,
\be
\eta_0=(72\pi)^\frac{1}{2}\frac{4\pi^2}{K^2(i)}\frac{e^{-\pi}}{(1+e^{-\pi})}
\ee
and
\be
\eta_1=\frac{\pi^2}{4K^2(i)}(2\pi)^\frac{1}{2}.
\ee
The aim of this rewriting of the mass gap is that, in this way, we are able to explicitly show the dependence of the energy scale $\mu$ and the 't Hooft coupling $\alpha_s$ in the mass gap. This makes it clear that Eqn.(\ref{eq:coup}) is a highly non-linear algebraic equation.

We observe that, due to the expansion with respect to $z=\Delta m^2/M^2 \ll 1$, our conclusion can be trusted only in the infrared limit. This is enough for a proof of confinement. In this limit, $\mu^2/M^2\rightarrow 0$, we can neglect the contribution coming from the exponential term in the mass gap and evaluate the corresponding beta function quite easily.

\medskip

\section{Non-local $\beta-$function}
\label{beta}

We derive the $\beta-$function by the following steps: 1) Firstly, we differentiate Eqn.(\ref{eq:coup}) with respect to $l=\ln(\mu^2/M^2)$. 2) We extract the factor $e^l$ from Eqn.(\ref{eq:coup}) and substitute it to the differential equation obtained in step 1). 3) Finally, we extract $d\alpha_s/dl$ from step 1). Within the limit of our approximations, \footnote{Note that the $\beta$-function we obtained is meaningful only in the IR-limit where our approximations hold. The effect of the non-locality appears to move the non-trivial fixed point to infinity, keeping confinement (no Landau-pole). In the UV-limit, we assume that asymptotic freedom holds, at least till the non-local mass scale that is assumed to be at very high energies.} we obtain for the first step 

\be
\beta_0\frac{d\alpha_s}{dl}\left(\frac{\gamma}{2} - (1 - \gamma)e^l\sqrt{\alpha_s}\eta_0\right) + \beta_0\alpha_s(l)\left(-(1 - \gamma)e^l\sqrt{\alpha_s(l)}\eta_0 - \frac{1}{2\sqrt{\alpha_s}}(1 - \gamma)e^l\eta_0
\frac{d\alpha_s}{dl}\right)=0,
\ee
with
\be
\beta_0=\frac{(N^2-1)^2}{2\pi N^2}.
\ee
From the second step, one has
\be
e^l=-\frac{\beta_0\alpha_s(l)\gamma + 2}{2\beta_0\alpha_s^\frac{3}{2}(l)\eta_0(\gamma - 1)}
\ee
In the given approximations, from the third step, the $\beta-$function is found to be a very simple form,
\be
\frac{d\alpha_s}{dl}=\beta(\alpha_s)=-2\frac{2+\beta_0\gamma\alpha_s}{6+\beta_0\gamma\alpha_s}\alpha_s.
\ee
The coupling runs to infinity in the infrared, signaling confinement, and runs to zero in the ultraviolet signaling the asymptotic freedom.

The equation we obtained is amenable to an exact solution. In order to avoid to clutter formulas, we show here the leading order of an asymptotic expansion holding in the IR limit. One has
\be
\label{eq:asIR}
\alpha_s(l)\sim \frac{(\beta_0^2\eta_0^2)^{-\frac{1}{3}}}{(1-\gamma)^\frac{2}{3}}\exp(-2l/3).
\ee
In the IR limit, $l\rightarrow -\infty$ and so, $\alpha_s(l)\rightarrow\infty$. In absence of a Landau ghost, this represents a confining theory. 

It is important to notice that, in this way, we have consistently extended the renormalization group to the non-local field theories. It is needed to emphasize that integration from 0 to infinity on the energy scale can take contributions also from the UV-limit, modifying the beta function. On the other side, in the UV-limit we have omitted important contributions in the derivation of the $\beta$-function and so, our results would need further improvement in this case. Anyway, we see that asymptotic freedom seems properly recovered but, due to our approximations, is mimicking the local case.

Finally, we can estimate the confinement scale by Eqn.(\ref{eq:asIR}), which is
\be
\mu^2_c\sim\left(\frac{\alpha_0}{\alpha_s}\right)^\frac{3}{2}M^2,
\ee
where we have set
\be
\alpha_0=\left[(1-\gamma)\beta_0\eta_0\right]^{-\frac{2}{3}}.
\ee
For SU(3), one has $\alpha_0\approx 0.2773$.



\begin{figure}[H]
\centering
\includegraphics{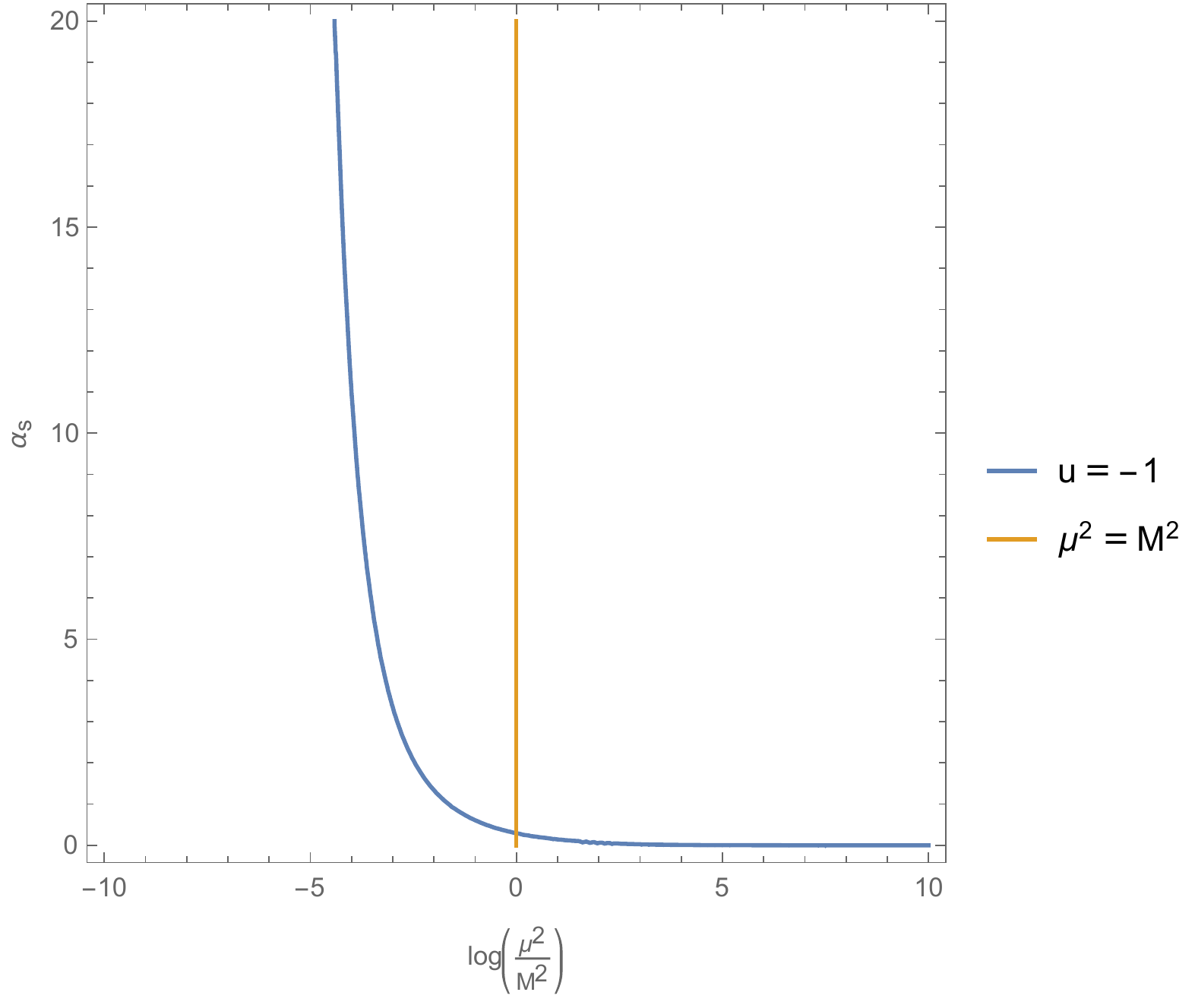}
\caption{\it Running coupling for non-local Yang-Mills theory with our approximations, showing the behavior in the infrared. $\mu $ is the RGE scale and M is the scale of non-locality. The coupling is seen to run to infinity smoothly at lower energies. Although one can see asymptotic freedom beyond the non-local scale, our solution is not applicable for this energy range.
}
\label{fig1}
\end{figure}

\smallskip

In Fig.\ref{fig1}, the running coupling is given with an indication of the starting of confinement limit at $\mu^2=M^2$. After that, the curve is seen to go to infinity steeply marking a confined regime of the theory, with the absence of Landau ghost.


\medskip

\section{Conclusions and Discussions}
\label{conc}


We investigated strongly coupled non-local gauge theory
in the 4-D in context of the confinement aspects of the theory. Alongside we compared the results with that of the local theory and discussed the implications regarding realistic QCD-like scenarios in the infra-red aspects. We presented the $\beta$-function of the theory and showed that the gauge condensation in the theory leads to dynamical generation of scales governed by the confinement scale and the scale of non-locality.  We summarize the main findings of our paper below:

\begin{enumerate}
    \item We derived the confinement conditions in the non-local gauge theories and showed that infinite-derivative non-local QFT in 4D provides confinement determined by the non-local scale M.
    \item We presented the Renormalization Group Equations for the non-local gauge theory in the strongly-coupled regimes and showed that the coupling runs to infinity in the low energy limit, without encountering the problem of Landau ghosts. The theory restores conformal invariance in the UV with strong coupling due to dilution of the mass gap arising from the non-local effects. 
    \item In the UV-limit, our conclusion can be trusted only until a certain energy scale due to the approximation made. In any case, the effects in the integral that defines the Kugo-Ojima $u$-function in Eqn.(\ref{eq:NL}) are summed till the non-locality scale yielding finite contributions also in the IR-limit and providing a proof of confinement, granted by the absence of the Landau ghost. The main difference with the local case is that the infrared fixed point is moved to infinity. 
\end{enumerate}

Our future studies will involve more detailed understanding of  confinement and $\beta$-function analysis in the framework of infinite-derivative non-local gravity theories \footnote{Whether such non-local factors may arise from any first-principle was recently discussed in Ref. \cite{Buoninfante:2021xxl}.} motivated from string field theory which provides UV-complete ghost-free, re-normalizable approach to quantum gravity which is free from cosmological \footnote{For inflationary cosmology \& predictions in the CMB, in non-local theories, see Refs.
\cite{Koshelev:2016vhi,SravanKumar:2018dlo,Koshelev:2020fok,Koshelev:2020xby,Koshelev:2020foq}.} and black-hole singularities \cite{Biswas:2011ar,Biswas:2013cha,Frolov:2015bia,Frolov:2015usa,Koshelev:2018hpt,Koshelev:2017bxd,Buoninfante:2018xiw,Cornell:2017irh,Buoninfante:2018rlq,Buoninfante:2018stt,Abel:2019zou,Buoninfante:2020ctr,Biswas:2005qr,Biswas:2006bs,Biswas:2010zk,Biswas:2012bp,Koshelev:2012qn,Koshelev:2018rau} while on the other hand leads to quadratic divergence-free, stable vacuum, no-Landau pole and confomally invariant QFT valid upto infinite energy scales \cite{Biswas:2014yia,Ghoshal:2017egr,Ghoshal:2018gpq,Buoninfante:2018gce,Ghoshal:2018gpq,Ghoshal:2020lfd,Frasca:2020jbe,Frasca:2021guf} \footnote{See Ref. \cite{conference} for recent conference on this topic.}.

\medskip 

\section{Acknowledgement}
\label{Asck}

This work is supported in part by the United States, Department of Energy Grant No. DE-SC0012447 (N.O.).

\medskip 

\section*{Appendix A: Dyson-Schwinger Equations \& Bender-Milton-Savage Technique}
\label{AppendixA}

In the following, we present the Bender-Milton-Savage technique \cite{Bender:1999ek}. This permits us to obtain the full hierarchy of Dyson-Schwinger equations in a PDE form.

Let us consider the partition function for a scalar field
\begin{equation}
    Z[j]=\int[D\phi]e^{iS(\phi)+i\int d^4xj(x)\phi(x)}.
\end{equation}
For the 1P-function we get
\be
\left\langle\frac{\delta S}{\delta\phi(x)}\right\rangle=j(x)
\ee
where
\be
\left\langle\ldots\right\rangle=\frac{\int[D\phi]\ldots e^{iS(\phi)+i\int d^4xj(x)\phi(x)}}{\int[D\phi]e^{iS(\phi)+i\int d^4xj(x)\phi(x)}}.
\ee
After that, we can complete the procedure by setting $j=0$. Next step is to derive the above equation for the 1P-function and dependent on $j$ to obtain the equation for the 2P-function. We emphasize that the definition of the nP-function is given by
\be
\langle\phi(x_1)\phi(x_2)\ldots\phi(x_n)\rangle=\frac{\delta^n\ln(Z[j])}{\delta j(x_1)\delta j(x_2)\ldots\delta j(x_n)}.
\ee
Therefore,
\be
\frac{\delta G_k(\ldots)}{\delta j(x)}=G_{k+1}(\ldots,x).
\ee

This means that, for a $\phi^4$ theory, one has
\be
S=\int d^4x\left[\frac{1}{2}(\partial\phi)^2-\frac{\lambda}{4}\phi^4\right],
\ee
so that,
\be
\label{eq:G_1}
\partial^2\langle\phi\rangle+\lambda\langle\phi^3(x)\rangle = j(x).
\ee
The following equation just holds
\be
Z[j]\partial^2G_1^{(j)}(x)+\lambda\langle\phi^3(x)\rangle = j(x).
\ee
By the definition of the 1P-function one gets
\be
Z[j]G_1^{(j)}(x)=\langle\phi(x)\rangle.
\ee
Now we derive with respect to $j(x)$ to obtain
\be
Z[j][G_1^{(j)}(x)]^2+Z[j]G_2^{(j)}(x,x)=\langle\phi^2(x)\rangle,
\ee
and after another derivation step it is
\be
Z[j][G_1^{(j)}(x)]^3+3Z[j]G_1^{(j)}(x)G_2(x,x)+Z[j]G_3^{(j)}(x,x,x)=\langle\phi^3(x)\rangle.
\ee
Inserting it into Eqn.(\ref{eq:G_1}) yields
\be
\label{eq:G1_j}
\partial^2G_1^{(j)}(x)+\lambda[G_1^{(j)}(x)]^3+3\lambda G_2^{(j)}(0)G_1^{(j)}(x)+G_3^{(j)}(0,0)=Z^{-1}[j]j(x)
\ee
We realize that, by the effect of renormalization, a mass term appeared.
We uncover here a term due to mass renormalization. Therefore, setting $j=0$, one gets the first Dyson-Schwinger equation into differential form
\be
\partial^2G_1(x)+\lambda[G_1(x)]^3+3\lambda G_2(0)G_1(x)+G_3(0,0)=0.
\ee

By deriving Eqn.(\ref{eq:G1_j}) again with respect to $j(y)$, we get
\be
\begin{split}
&\partial^2G_2^{(j)}(x,y)+3\lambda[G_1^{(j)}(x)]^2G_2^{(j)}(x,y)+
\nonumber \\
&3\lambda G_3^{(j)}(x,x,y)G_1^{(j)}(x)
+3\lambda G_2^{(j)}(x,x)G_2^{(j)}(x,y)
+G_4^{(j)}(x,x,x,y)=\nonumber \\
&Z^{-1}[j]\delta^4(x-y)+j(x)\frac{\delta}{\delta j(y)}(Z^{-1}[j]).
    \end{split}
\ee
Inserting $j=0$, the equation for the 2P-function takes the form
\be
\partial^2G_2(x,y)+3\lambda[G_1(x)]^2G_2(x,y)+
3\lambda G_3(0,y)G_1(x)
+3\lambda G_2(0)G_2(x,y)
+G_4(0,0,y)=
\delta^4(x-y).
\ee
This procedure can be iterated to any wished order providing all the hierarchy of Dyson-Schwinger equations in PDE form.


\section*{Appendix B: Dyson-Schwinger equations for 1P- and 2P-functions}
\label{AppendixB}

In this appendix, we derive the Dyson-Schwinger equations for the Yang-Mills field for the 1P- and 2P-functions.

For the 1P-functions, after averaging the equations of motion, we get
\begin{eqnarray}
\Box G_{1\mu}^{(j)a}+gf^{abc}
e^{-\frac{1}{2}f(\Box)}\left\langle\partial_\nu\left[e^{\frac{1}{2}f(\Box)}{\bar A}^{b}_\mu
e^{\frac{1}{2}f(\Box)}A^{c\nu}\right]\right\rangle+&& \nonumber \\
gf^{abc}e^{-\frac{1}{2}f(\Box)}\left\langle\left[ e^{\frac{1}{2}f(\Box)}A^{b\nu}
e^{\frac{1}{2}f(\Box)}(\partial_\mu A^{c}_\nu-\partial_\nu A^{c}_\mu)\right]\right\rangle&& \nonumber \\
g^2f^{abc}f^{cde}e^{-\frac{1}{2}f(\Box)}
\left\langle\left[e^{\frac{1}{2}f(\Box)}A^{b\nu}e^{\frac{1}{2}f(\Box)}A^{d}_\nu
e^{\frac{1}{2}f(\Box)}A^{e}_\mu\right]\right\rangle+&& \nonumber \\
+gf^{abc}e^{\frac{1}{2}f(\Box)}\left\langle \bar{c}^b\partial_\mu c^c\right\rangle&=&e^{\frac{1}{2}f(\Box)}j^a_\mu,
\end{eqnarray}
and for the ghost
\be
-\Box P_1^{(\eta)a} +gf^{abc}\left\langle\left(e^{\frac{1}{2}f(\Box)}A_\mu^c\right)\partial^\mu c^b\right\rangle=e^{\frac{1}{2}f(\Box)}\eta^a.
\ee
For our aims, we introduced the following 1P-functions
\bea
\label{eq:defs}
G_{1\mu}^{(j)a}(x)&=&Z^{-1}\langle A_\mu^a(x)\rangle \nonumber \\
P_1^{(\eta)a}(x)=&=&Z^{-1}\langle c^a(x)\rangle.
\eea
The same should hold for ${\bar c}^a$ yielding ${\bar P}_1^{(\eta)a}(x)$. 
In order to evaluate the averages we consider the above definitions rewritten as
\bea
Z[j,\eta,{\bar\eta}]e^{\frac{1}{2}f(\Box)}G_{1\mu}^{(j)a}(x)&=&\langle e^{\frac{1}{2}f(\Box)}A_\mu^a(x)\rangle \nonumber \\
Z[j,\eta,{\bar\eta}]P_1^{(\eta)a}(x)&=&\langle c^a(x)\rangle.
\eea
The apexes $(j)$ and $(\eta)$ are there to remember the explicit dependence on the currents. Let us derive one time with respect to $j(x)$ on the first equation to get
\be
\label{eq:f1}
Ze^{\frac{1}{2}f(\Box)}G_{2\mu\nu}^{(j)ab}(x,x)+
Ze^{\frac{1}{2}f(\Box)}G_{1\mu}^{(j)a}(x)e^{\frac{1}{2}f(\Box)}G_{1\nu}^{(j)b}(x)=
\langle e^{\frac{1}{2}f(\Box)}A_\mu^a(x)e^{\frac{1}{2}f(\Box)}A_\nu^b(x)\rangle.
\ee
We apply the space-time derivative $\partial^\nu$ obtaining
\be
Ze^{\frac{1}{2}f(\Box)}\partial^\nu G_{2\mu\nu}^{(j)ab}(x,x)+
Ze^{\frac{1}{2}f(\Box)}\partial^\nu G_{1\mu}^{(j)a}(x)e^{\frac{1}{2}f(\Box)}G_{1\nu}^{(j)b}(x)=
\langle e^{\frac{1}{2}f(\Box)}\partial^\nu A_\mu^a(x)e^{\frac{1}{2}f(\Box)}A_\nu^b(x)\rangle.
\ee
We further derive Eqn.(\ref{eq:f1}) with respect to $j^{c\nu}$ to get
\bea
Ze^{\frac{1}{2}f(\Box)}G_{2\mu\nu}^{(j)ab}(x,x)e^{\frac{1}{2}f(\Box)}G_1^{(j)\nu c}(x)
+Ze^{\frac{1}{2}f(\Box)}G_{3\mu\nu}^{(j)abc\nu}(x,x,x)+
\nonumber \\
Ze^{\frac{1}{2}f(\Box)}G_{1\mu}^{(j)a}(x)e^{\frac{1}{2}f(\Box)}G_{1\nu}^{(j)b}(x)e^{\frac{1}{2}f(\Box)}G_1^{(j)\nu c}(x)+Ze^{\frac{1}{2}f(\Box)}G_{2\mu}^{(j)ac\nu}(x)e^{\frac{1}{2}f(\Box)}G_{1\nu}^{(j)b}+\nonumber \\
Ze^{\frac{1}{2}f(\Box)}G_{2\nu}^{(j)bc\nu}(x)e^{\frac{1}{2}f(\Box)}G_{1\mu}^{(j)a}(x)=
\langle e^{\frac{1}{2}f(\Box)}A_\mu^a(x)e^{\frac{1}{2}f(\Box)}A_\nu^b(x)e^{\frac{1}{2}f(\Box)}A^{c\nu}(x)\rangle,
\eea
and we need to do the same for the ghost field. From Eqn.(\ref{eq:defs}) we write
\be
\label{eq:P1}
Z[j,\eta,\bar\eta]P_1^{(\eta)a}(x)=\langle c^a(x)\rangle.
\ee
After deriving with respect to $\partial_\mu$ and then with respect to $\bar\eta$, one has
\be
Z{\bar P}_1^{(\eta)b}(x)e^{\frac{1}{2}f(\Box)}\partial^\mu P_1^{(\eta)a}(x)+Z\partial^\mu K_2^{(\eta)ab}(x,x)=
\langle {\bar c}^b\partial^\mu c^a(x)\rangle.
\ee
We have introduced a new 2P-function defined as 
\be
K_2^{(\eta)ab}(x,y)=\frac{1}{Z}\frac{\delta P_1^{(\eta)a}(x)}{\delta \eta^b(y)},
\ee
and the other 2P-function
\be
J_{2\mu}^{(\eta,j)ab}(x,y)=\frac{1}{Z}\frac{\delta P_1^{(\eta)a}(x)}{\delta j^{b\mu}(y)}.
\ee
So, by deriving Eqn.(\ref{eq:P1}) with respect to $j^{b\mu}(x)$, the result is
\be
Ze^{\frac{1}{2}f(\Box)}G_{1\mu}^{(j)b}(x)\partial^\mu P_1^{(\eta)a}(x)+Z\partial^\mu J_{2\mu}^{(\eta,j)ab}(x,x)=
\langle A^b_\mu(x)\partial^\mu c^a(x)\rangle.
\ee
Collecting everything together, one has
\begin{eqnarray}
\label{eq:G1j}
\Box G_{1\mu}^{(j)a}+gf^{abc}
e^{-\frac{1}{2}f(\Box)}\partial^\nu\left[
e^{\frac{1}{2}f(\Box)}G_{2\mu\nu}^{(j)bc}(x,x)+
e^{\frac{1}{2}f(\Box)}G_{1\mu}^{(j)b}(x)e^{\frac{1}{2}f(\Box)}G_{1\nu}^{(j)c}(x)
\right]-&& \nonumber \\
gf^{abc}e^{-\frac{1}{2}f(\Box)}
\left[ 
e^{\frac{1}{2}f(\Box)}\partial^\nu G_{2\mu\nu}^{(j)bc}(x,x)+
e^{\frac{1}{2}f(\Box)}\partial^\nu G_{1\mu}^{(j)b}(x)e^{\frac{1}{2}f(\Box)}G_{1\nu}^{(j)c}(x)
\right]-
&& \nonumber \\
gf^{abc}e^{-\frac{1}{2}f(\Box)}
\left[ 
e^{\frac{1}{2}f(\Box)}\partial_\mu G_{2\nu}^{(j)bc\nu}(x,x)+
e^{\frac{1}{2}f(\Box)}\partial_\mu G_{1\nu}^{(j)b}(x)e^{\frac{1}{2}f(\Box)}G_{1}^{(j)c\nu}(x)
\right]+
&& \nonumber \\
g^2f^{abc}f^{cde}e^{-\frac{1}{2}f(\Box)}
\left[
e^{\frac{1}{2}f(\Box)}G_{2\mu\nu}^{(j)bd}(x,x)e^{\frac{1}{2}f(\Box)}G_1^{(j)\nu e}(x)
+e^{\frac{1}{2}f(\Box)}\partial^\nu G_{3\mu\nu}^{(j)bde\nu}(x,x,x)+
\right.
\nonumber \\
e^{\frac{1}{2}f(\Box)}G_{1\mu}^{(j)b}(x)e^{\frac{1}{2}f(\Box)}G_{1\nu}^{(j)d}(x)e^{\frac{1}{2}f(\Box)}G_1^{(j)\nu e}(x)+e^{\frac{1}{2}f(\Box)}G_{2\mu}^{(j)be\nu}(x,x)e^{\frac{1}{2}f(\Box)}G_{1\nu}^{(j)d}(x)+\nonumber \\
\left.
e^{\frac{1}{2}f(\Box)}G_{2\nu}^{(j)de\nu}(x,x)e^{\frac{1}{2}f(\Box)}G_{1\mu}^{(j)b}(x)
\right]-&& \nonumber \\
gf^{abc}e^{\frac{1}{2}f(\Box)}
\left\{
{\bar P}_1^{(\eta)b}(x)e^{\frac{1}{2}f(\Box)}\left[\partial_\mu P_1^{(\eta)c}(x)\right]+\partial_\mu\left[K_2^{(\eta)bc}(x,x)\right]
\right\}
&=& \nonumber \\
e^{\frac{1}{2}f(\Box)}j^a_\mu,&&
\end{eqnarray}
The equation of the local theory given in \cite{Frasca:2015yva} are easily obtained by setting the non-locality factor to 1, corresponding to the local limit $M \rightarrow \infty$
For the ghost field, it is
\be
-\Box P_1^{(\eta)c} 
-gf^{abc}e^{\frac{1}{2}f(\Box)}G_{1\mu}^{(j)a}(x)\partial^\mu P_1^{(\eta)b}(x)-gf^{abc}\partial^\mu J_{2\mu}^{(\eta,j)ab}(x,x)
=e^{\frac{1}{2}f(\Box)}\eta^c.
\ee
After setting all the current to zero, the Dyson-Schwinger equations for the 1P-functions are obtained in the form given in the main text.

From Eqn.(\ref{eq:G1j}), we derive it with respect to $j^{\lambda h}(y)$ obtaining
\begin{eqnarray}
\label{eq:G2j}
\Box G_{2\mu\lambda}^{(j)ah}(x,y)+gf^{abc}
e^{-\frac{1}{2}f(\Box)}\partial^\nu\left[
e^{\frac{1}{2}f(\Box)}G_{3\mu\nu\lambda}^{(j)bch}(x,x,y)+
e^{\frac{1}{2}f(\Box)}G_{2\mu\lambda}^{(j)bh}(x,y)\times
\right.&&
\nonumber \\
\left.
e^{\frac{1}{2}f(\Box)}G_{1\nu}^{(j)c}(x)+
+e^{\frac{1}{2}f(\Box)}G_{1\mu}^{(j)b}(x)
e^{\frac{1}{2}f(\Box)}G_{2\nu\lambda}^{(j)ch}(x)
\right]-&& \nonumber \\
gf^{abc}e^{-\frac{1}{2}f(\Box)}
\left[ 
e^{\frac{1}{2}f(\Box)}\partial^\nu G_{2\mu\nu\lambda}^{(j)bch}(x,x,y)+
e^{\frac{1}{2}f(\Box)}\partial^\nu G_{2\mu\lambda}^{(j)bh}(x,y)e^{\frac{1}{2}f(\Box)}G_{1\nu}^{(j)c}(x)+
\right.
&& \nonumber \\
\left.
e^{\frac{1}{2}f(\Box)}\partial^\nu G_{1\mu}^{(j)b}(x)e^{\frac{1}{2}f(\Box)}G_{2\nu\lambda}^{(j)ch}(x,y)
\right]-
&& \nonumber \\
gf^{abc}e^{-\frac{1}{2}f(\Box)}
\left[
e^{\frac{1}{2}f(\Box)}\partial_\mu G_{3\nu\lambda}^{(j)bch\nu}(x,x,y)+
e^{\frac{1}{2}f(\Box)}\partial_\mu G_{2\nu\lambda}^{(j)bh}(x,y)e^{\frac{1}{2}f(\Box)}G_{1}^{(j)c\nu}(x)+
\right.&&
\nonumber \\
\left.
e^{\frac{1}{2}f(\Box)}\partial_\mu G_{1\nu}^{(j)b}(x)e^{\frac{1}{2}f(\Box)}G_{2\lambda}^{(j)ch\nu}(x,y)
\right]+
&& \nonumber \\
g^2f^{abc}f^{cde}e^{-\frac{1}{2}f(\Box)}
\left[
e^{\frac{1}{2}f(\Box)}G_{3\mu\nu\lambda}^{(j)bdh}(x,x,y)e^{\frac{1}{2}f(\Box)}G_1^{(j)\nu e}(x)+
\right.&&
\nonumber \\
e^{\frac{1}{2}f(\Box)}G_{2\mu\nu}^{(j)bd}(x,x)e^{\frac{1}{2}f(\Box)}G_{2\lambda}^{(j)\nu eh}(x,y)
+e^{\frac{1}{2}f(\Box)}\partial^\nu G_{4\mu\nu\lambda}^{(j)bdeh\nu}(x,x,x,y)+
\nonumber \\
e^{\frac{1}{2}f(\Box)}G_{2\mu\lambda}^{(j)bh}(x,y)e^{\frac{1}{2}f(\Box)}G_{1\nu}^{(j)d}(x)e^{\frac{1}{2}f(\Box)}G_1^{(j)\nu
e}(x)+&&
\nonumber \\
e^{\frac{1}{2}f(\Box)}G_{1\mu}^{(j)b}(x)e^{\frac{1}{2}f(\Box)}G_{2\nu\lambda}^{(j)dh}(x,y)e^{\frac{1}{2}f(\Box)}G_1^{(j)\nu
e}(x)+&&
\nonumber \\
e^{\frac{1}{2}f(\Box)}G_{1\mu}^{(j)b}(x)e^{\frac{1}{2}f(\Box)}G_{1\nu}^{(j)d}(x)e^{\frac{1}{2}f(\Box)}G_{2\lambda}^{(j)\nu
eh}(x,y)+&&
\nonumber \\
e^{\frac{1}{2}f(\Box)}G_{3\mu\lambda}^{(j)beh\nu}(x,x,y)e^{\frac{1}{2}f(\Box)}G_{1\nu}^{(j)d}(x)+&& 
\nonumber \\
e^{\frac{1}{2}f(\Box)}G_{2\mu}^{(j)be\nu}(x,x)e^{\frac{1}{2}f(\Box)}G_{2\nu\lambda}^{(j)dh}(x,y)+&&
\nonumber \\
\left.
e^{\frac{1}{2}f(\Box)}G_{3\nu\lambda}^{(j)deh\nu}(x,x,y)e^{\frac{1}{2}f(\Box)}G_{1\mu}^{(j)b}(x)
+e^{\frac{1}{2}f(\Box)}G_{2\nu}^{(j)de\nu}(x,x)e^{\frac{1}{2}f(\Box)}G_{2\mu\lambda}^{(j)bh}(x,y)
\right]-&& \nonumber \\
gf^{abc}e^{\frac{1}{2}f(\Box)}
\left\{
{\bar J}_{2\lambda}^{(\eta,j)bh}(x,y)e^{\frac{1}{2}f(\Box)}\left[\partial_\mu P_1^{(\eta)c}(x)\right]\right.+&& \nonumber \\
\left.{\bar P}_1^{(\eta)b}(x)e^{\frac{1}{2}f(\Box)}\left[\partial_\mu J_{2\lambda}^{(\eta)ch}(x,y)\right]
+\partial_\mu\left[W_{3\lambda}^{(\eta,j)bch}(x,x,y)\right]
\right\}=&& \nonumber \\
e^{\frac{1}{2}f(\Box)}\delta^{ah}\eta_{\mu\lambda}\delta^4(x-y),&&
\end{eqnarray}
after the introduction of the 3P-function
\be
W_{3\lambda}^{(\eta,j)abc}(x,y,z)=Z^{-1}\frac{\delta K_2^{(\eta)ab}(x,y)}{\delta j^{\lambda c}(z)}.
\ee
Similarly, starting from the 1P-function for the ghost and deriving it with respect to $\eta^{h}(y)$, we get
\bea
&-\Box K_2^{(\eta)ch}(x,y)
-ige^{\frac{1}{2}f^{abc}f(\Box)}L_{2\mu}^{(\eta,j)ah}(x,y)\partial^\mu P_1^{(\eta)b}(x)\nonumber \\
&-igf^{abc}e^{\frac{1}{2}f(\Box)}G_{1\mu}^{(j)a}(x)\partial^\mu K_2^{(\eta)bh}(x,y)
-igf^{abc}\partial^\mu W_{3\mu}^{(\eta,j)abh}(x,x,y) \nonumber \\
&=e^{\frac{1}{2}f(\Box)}\delta^{ch}\delta^4(x-y).
\eea
We have introduced the 2P-function
\be
L_{2\mu}^{(\eta,j)ab}(x,y)=\frac{\delta G_1^{(j)a}(x)}{\delta \eta^b(y)}.
\ee
Deriving with respect to $j^{h\nu}(y)$, one has the equation for $J_2$ in the form
\bea
&-\Box J_2^{(\eta)ch\nu}(x,y) 
-igf^{abc}e^{\frac{1}{2}f(\Box)}G_{2\mu\nu}^{(j)ah}(x,y)\partial^\mu P_1^{(\eta)b}(x) \nonumber \\
&-igf^{abc}e^{\frac{1}{2}f(\Box)}G_{1\mu}^{(j)a}(x)\partial^\mu J_2^{(\eta,j)bh\nu}(x,y) \nonumber \\
&-igf^{abc}\partial^\mu J_{3\mu}^{(\eta,j)abh}(x,x,y)=0,
\eea
with the introduction of the 3P-function
\be
J_{3\mu}^{(\eta,j)abc}(x,y,z)=\frac{\delta J_{2\mu}^{(\eta,j)ab}(x,y)}{\delta j^{c\mu}(z)}.
\ee
We can recover the equations in the main text after setting all the currents to zero.

\section*{Appendix C: Confinement in local Yang-Mills theory}
\label{AppendixC}

The approach given in the main text is straightforwardly obtained by the analysis performed in Ref. \cite{Chaichian:2018cyv}. The $u$ function can be obtained by observing that
\begin{eqnarray}
\int d^4xe^{ipx}\langle D_\mu\bar{c}^{a} (x),D_\nu c^{b} (0) \rangle&=&-\delta^{ab}\frac{p_\mu p_\nu}{k^2}\\
&+&\frac{(N^2-1)^2}{2N}g^2\delta^{ab}
\left(\delta_{\mu\nu} - \frac{p_{\mu} p_{\nu}}{p^{2}}\right)
\int\frac{d^4p'}{(2\pi)^4}K_2(p-p')G_2(p').\nonumber
\end{eqnarray}
For local Yang-Mills theory, the propagators take the form
\be
K_2(p)=-\frac{1}{p^2+i\epsilon}
\ee
for the ghost field and
\be
G_2(p)=\frac{\pi^3}{4K^3(i)}
	\sum_{n=0}^\infty\frac{e^{-(n+\frac{1}{2})\pi}}{1+e^{-(2n+1)\pi}}(2n+1)^2\frac{1}{p^2-m_n^2+i\epsilon}
\ee
for the gauge field, with the mass spectrum
\be
m_n=(2n+1)\frac{\pi}{2K(i)}\left(\frac{Ng^2}{2}\right)^\frac{1}{4}\mu,
\ee
where $K(i)$ is the complete elliptical integral of the first kind and $\mu$ is one of the integration constants of the theory. This is an approximate solution as we have neglected any mass shift arising from renormalization. This will yield for the confinement condition by Kugo-Ojima
\be
\label{eq:NL}
u(0)=-\frac{(N^2-1)^2}{2N}g^2
\int\frac{d^4p}{(2\pi)^4}\frac{1}{p^2+i\epsilon}\frac{\pi^3}{4K^3(-1)}
	\sum_{n=0}^\infty\frac{e^{-(n+\frac{1}{2})\pi}}{1+e^{-(2n+1)\pi}}(2n+1)^2\frac{1}{p^2-m_n^2+i\epsilon}=-1.
\ee
This integral can be performed by a continuation in the complex plane, moving to Euclidean. It is divergent and can be evaluated e.g. by dimensional regularization. This will yield the beta function in closed form for the theory, holding both in the UV and IR, as
\be
\beta_{YM}=-\beta_0\frac{\alpha_s^2}{1-\frac{1}{2}\beta_0\alpha_s},
\ee
with $\beta_0=(N^2-1)^2/8\pi N$. This beta function grants confinement in the IR having a non-trivial finite fixed point. In the UV we recover the asymptotic freedom as expected.

\bibliographystyle{unsrt}

\end{document}